\documentclass[twocolumn]{aastex631}

\newcommand{\nfig}[1]{Figure~\ref{#1}}
\newcommand{\speed}[1]{#1 km~s${}^{-1}$}
\newcommand{\accel}[1]{#1 km~s${}^{-2}$}
\received{Jun 15, 2021}
\revised{Aug 30, 2021}
\accepted{\today}
\submitjournal{ApJ}

\shorttitle{Sympathetic Filament Eruptions}
\shortauthors{Zhou et al.}

\begin{document}
\title{Sympathetic Filament Eruptions within a Fan-spine Magnetic System}
\author{Chengrui Zhou}
\affiliation{Yunnan Observatories, Chinese Academy of Sciences, Kunming 650216, China}
\affiliation{State Key Laboratory of Space Weather, Chinese Academy of Sciences, Beijing 100190, China}
\affiliation{University of Chinese Academy of Sciences, Beijing, 100049, China}
\author[0000-0001-9493-4418]{Yuandeng Shen}
\affiliation{Yunnan Observatories, Chinese Academy of Sciences, Kunming 650216, China}
\affiliation{State Key Laboratory of Space Weather, Chinese Academy of Sciences, Beijing 100190, China}
\affiliation{Key Laboratory of Solar Activity, National Astronomical Observatories, Chinese Academy of Sciences, Beijing 100012, China}
\affiliation{University of Chinese Academy of Sciences, Beijing, 100049, China}
\author{Xinping Zhou}
\affiliation{Yunnan Observatories, Chinese Academy of Sciences, Kunming 650216, China}
\affiliation{University of Chinese Academy of Sciences, Beijing, 100049, China}
\author{Zehao Tang}
\affiliation{Yunnan Observatories, Chinese Academy of Sciences, Kunming 650216, China}
\affiliation{University of Chinese Academy of Sciences, Beijing, 100049, China}
\author{Yadan Duan}
\affiliation{Yunnan Observatories, Chinese Academy of Sciences, Kunming 650216, China}
\affiliation{University of Chinese Academy of Sciences, Beijing, 100049, China}
\author{Song Tan}
\affiliation{Yunnan Observatories, Chinese Academy of Sciences, Kunming 650216, China}
\affiliation{University of Chinese Academy of Sciences, Beijing, 100049, China}
\correspondingauthor{Yuandeng Shen}
\email{ydshen@ynao.ac.cn}

\begin{abstract}
It is unclear whether successive filament eruptions at different sites within a short time interval are physically connected or not. Here, we present the observations of the successive eruptions of a small and a large filament in a tripolar magnetic field region whose coronal magnetic field showed as a fan-spine magnetic system. By analyzing the multi-wavelength observations taken by the Solar Dynamic Observatory (SDO) and the extrapolated three-dimensional coronal magnetic field, we find that the two filaments resided respectively in the two lobes that make up the inner fan structure of the fan-spine magnetic system. In addition, a small fan-spine system was also revealed by the squashing factor Q map, which located in the east lobe of the fan structure of the large fan-spine system. The eruption of the small filament was a failed filament eruption, which did not cause any coronal mass ejection (CME) except for three flare ribbons and two post-flare-loop systems connecting the three magnetic polarities. The eruption of the large filament not only caused similar post-flare-loop systems and flare ribbons as observed in the small filament eruption, but also a large-scale CME. Based on our analysis results, we conclude that the two successive filament eruptions were physically connected, in which the topology change caused by the small filament eruption is thought to be the physical linkage. In addition, the eruption of the small fan-spine structure further accelerated the instability and violent eruption of the large filament.  
\end{abstract}
\keywords{Sun: activity --- Sun: flares --- Sun: filaments --- Sun: chromosphere --- Sun: magnetic fields}

\section{Introduction}\label{intro}
Solar filaments (or prominences) are one of the basic building blocks of the solar atmosphere, their eruption into the solar-terrestrial space can cause disastrous space weather \citep{2011LRSP....8....1C,2013ScChD..56.1091W}. High spatiotemporal observations showed that the main body of a filament is composed of many thin threads carrying many dynamical mass flows, transverse waves, and Kelvin-Helmholtz instabilities \citep[e.g.,][]{2015ApJ...814L..17S,2018ApJ...863..192L,2021A&A...647A.112Z,2007Sci...318.1577O}. The supporting magnetic structure of a filament could be dips in sheared arcades \citep{kip57,lite05,Mackay2010,Aulanier1998a} or twisted magnetic flux ropes (MFRs) \citep{rust96,zhangj12,Zhou2017}, while its mass can be originated from direct injection of chromospheric cold material \citep[e.g.,][]{1999ApJ...520L..71W,2003ApJ...584.1084C,shen19a}, the condensation of hot coronal plasma due to thermal instability \citep{hil74,spark90,xia2011, xia2012, xia2014}, or a combination of the two \citep{2021arXiv210413546H}. In filament physics, there are many questions yet to be resolved. Besides the formation and stability, the eruption mechanism is still not understood completely. Generally, filament eruptions can be divided into three types, i.e., failed, partial and full eruptions \citep[][]{2001ApJ...549.1221G,ji03,2011RAA....11..594S,shen12b}. However, what determines the final result of a filament eruption is also unclear. In previous studies, many theoretical models have been proposed to interpret successful filament eruptions, and most of them take magnetic reconnection as the basic trigger and driving mechanism \citep[e.g.,][]{chen00,lin&forbes2000,moo01,ant99}. In recent years, ideal magnetohydrodynamic (MHD) instabilities such as kink instability and torus instability are also thought to be important for the onset of filament eruptions \citep{ji03,kumar11,joshi13,joshi14,xuh20}. 

Observations indicate that solar eruptions are very complicated, which represent the large-scale  rearrangement of mass and magnetic field and energy conversion, and are accompanied by spectacular phenomena such as flares \citep{2011LRSP....8....6S}, CMEs \citep[e.g.,][]{2011LRSP....8....1C,Lynch08,Lynch13,2019ApJ...887..118C}, filament eruptions \citep[e.g.,][]{bi14,Panesar17,2020ApJ...902....8C}, jets \citep[e.g.,][]{shi07,shen12b,shen17,shen19a,str17,shen21}, and magnetohydrodynamic waves \citep{shen18,miao19}. High-resolution observations in recent years showed that the eruption of many small-scale solar eruptions, such as mini-filament eruptions and solar jets, resemble their large-scale counterparts, indicating the possible scale invariance of solar eruptions \citep[e.g.,][]{shen12b,wang12,Lim2016,Panesar18,Li2018,yang18,shen19a,duan2019}. Normally, a solar eruption occurs independently. However, for some successive events occurring within a short time interval in the same complicated active region or at different active regions, they may physically connect to each other, i.e., the sympathetic eruptions in which the occurrence of one event makes another one at different places \citep{1998ApJ...509..448W}. The physical linkages in sympathetic eruptions can be the disruption of large-scale convective motions \citep{bum1993}, the interaction between different magnetic flux systems \citep{peng07,jiang08,tang21}, the large-scale waves \citep{shen14,2014ApJ...795..130S}, and reduction of confining magnetic fields caused by the earlier eruptions \citep[e.g.,][]{shen12a,hou20,yang12,jiang11}. In addition, \cite{sch11} proposed that long-range flares and eruptions could be connected by a system of separatrices, separators, and quasi-separatrix layers \citep{2012ApJ...759...70T}. Usually, the removal of the open magnetic fields or the external disturbances were thought as a key signature indicative for the occurrence of sympathetic eruptions. \citet{jiang11} observed a filament eruption which subsequently caused the removal of overlying loops of two adjacent filaments and resulted in their eruption, during which the dimming regions could be a good agent for linking consecutive solar eruptions. \citet{hou20} investigated the consecutive eruption of two nearby filaments, in which the authors found that the first filament eruption pushed its overlying loops to reconnect with the overlying loop of the other filament. Therefore, the filament became unstable and then erupted due to the reduction of the overlying magnetic fields by the external magnetic reconnection \citep{2020ApJ...892...79S}. \citet{shen12a} reported two consecutive filament eruptions in a breakout magnetic configuration, including a partial filament eruption and a full successful filament eruption. They firstly proposed that the so-called  magnetic implosion mechanism \citep{hu00} can be the physical linkage for the observed sympathetic filament eruptions, in which the external null point reconnection played a vital role in the initiation of the event. Sympathetic filament eruptions have been studied in several numerical simulation studies. \citet{Lynch13} performed a 2.5D MHD simulation which revealed two breakout CMEs in a sympathetic filament eruption from a pseudostreamer configuration, its eruption paired with the current sheets developed by the external and the internal reconnections. \citet{tor11} numerically studied the eruption mechanism of the consecutive eruptions of three nearby filaments, in which two filaments were confined by a pseudostreamer, while the other one was next to it. The eruption of the outside filament resulted in the consecutive eruption of the two filaments inside the pseudostreamer. The simulation suggested that the key of the sympathetic filament eruption was the consecutive consumption of the filaments' overlying magnetic fluxes by the external reconnections. All these observational and numerical studies suggest that the large-scale structural properties of the coronal magnetic field are important for the occurrence of solar sympathetic eruptions, and the physical linkages for different types of eruptions can be diversified. Particularly, we will focus on the magnetic topology change as the mechanism for sympathetic filament eruptions in regions with multiple magnetic polarities.

As firstly proposed by \cite{hu00}, a magnetic implosion occurs following the basic law that transient events such as flares and CMEs should release free magnetic energy into the low plasma $\beta$ corona, and the reduction of magnetic energy inevitably result in the decreasing of the magnetic pressure around the energy releasing regions. The magnetic implosion can be indirectly estimated by the contraction of coronal loops, and those have been observed in several events \citep[e.g.,][]{lw2010,gos12,liuliu12}. \citet{wang16} reported a well observed contraction of the imploding loops, and the implosion was actually took place during the eruptive phase of the associated filament. In order to refrain from the projection effect, \citet{pit16} and \citet{wangjun18} analyzed the contraction of coronal loops from two different observing angles, their study confirmed that the contraction of corona loops is indeed associated with flares. \citet{sim13} estimated the energy redistribution and implosive motions in a solar flare, and they found that the contraction happened during the flare's impulsive phase, and the contraction rate was closely associated with the intensity of the hard X-ray and microwave emissions. Observations showed that the occurrence of magnetic implosion is commonly accompanied by the oscillation of coronal loops (or part loops) besides loop contraction \citep[e.g.,][]{lw2010,gos12,Sun12,sim13,rus15}. However, \citet{rus15} proposed that contractions and oscillations of coronal loops can occur in a single response to the occurrence of magnetic implosion, and they argued that coronal magnetic implosion is a newly identified excitation mechanism for transverse loops oscillation. The relation between the converging motions of conjugate loop's footpoints and the descending motion of loop top sources at the beginning of solar flares were also explained as the results of magnetic implosions \cite[e.g.,][]{2003ApJ...596L.251S,2004ApJ...607L..55J,2006A&A...446..675V}. Recently, coronal magnetic implosion was also used to explain the abrupt reserved in the rotation and some penumbral dynamics of sunspots \citep{2016NatCo...713798B,2019ApJ...874..134X}, and the physical linkage of sympathetic filament eruptions within the framework of magnetic breakout configuration \citep{shen12b}.

Solar eruptions are more liable to occur in complicated active regions with sufficient magnetic free energy \citep{munro1979,zhang01}. A tripolar magnetic region often manifests as a central parasitic magnetic field encompassing by the opposite polarity, and the corresponding overlying coronal magnetic field shows as a fan-spine magnetic configuration consisting of a dome-shaped fan structure, a null point, and inner and outer spines \citep{lau90,Masson09,shen19b}. The far end of the outer spine connects to a remote magnetic field region with the same polarity as the inner parasitic magnetic field. Therefore, to a certain degree, a complete fan-spine magnetic structure is in fact a magnetic breakout topology in nature \citep{ant99,shen12b,2017Natur.544..452W,shen21}. Solar eruptions in fan-spine magnetic structures have been studied by many researchers \citep{lau90,tor09,tor10,Masson09,par09,wyp18}, and they are found to be in favor of the occurrence of solar jets  \citep{shen19b,shen21}. Previous studies suggested that the dome-shaped fan structure can be characterized by the high degree of squashing factor Q, and it is actually a separatrix layer in favor of the occurrence of magnetic reconnection. When magnetic reconnection happens in the fan-separatrix, it changes the connectivities of the neighboring magnetic field lines, and leads to a flare. Generally, flares associated with the eruptions in fan-spine structures often consist of three flare ribbons: an inner bright point surrounded by a circular ribbon, and a remote brightening \citep{wang12,shen19b,Lih17}. The formation of flare ribbons corresponds to the movement of the footprints of separatrices or quasi-separatrices (QSLs); therefore, the flare ribbons show the magnetic reconnection process of flares \citep{priest1996}. In many cases, the eruption of fan-spine structures involves the eruption of mini-filaments in the fan structures, and the unstable of the fan-spine structures were believed to result from the reconnection around the null point, with a great deal of confined arcades consumed \citep{sun13,shen19b,yang20}. According to standard filament eruption models \citep[e.g.,][]{lin&forbes2000}, magnetic reconnection will take place below the rising filament, and cause two parallel conjugated flare ribbons. Within the framework of fan-spine magnetic systems, the rising filament also pushes its confining field lines to reconnect with the reconnection favorable field lines around the null point. Therefore, a paired magnetic reconnections can be expected in such eruptions. Normally, the reconnection below the rising filament is called internal reconnection, while that around the null point is called external reconnection. The external reconnection can occur before or after the internal one, and in different cases, one can observe different eruption characteristics \citep{joshi15,zhang15,shen19b}. Eruptions in fan-spine systems in or around active regions are evidenced to be hard to produce successful CMEs in the interplanetary space \citep[e.g.,][]{Lih19,shen19b,Yangs2020}. So far only a few observations showed the production of CMEs from fan-spine magnetic systems, and those events were all associated with energetic large flares \citep[e.g.,][]{Lih19,2020ApJ...899...34L}. The large-scale pseudostreamers can be viewed as the larger counterpart of fan-spine systems, nothing but with their outer spine extending to the interplanetary space. It seems that eruptions in pseudostreamers are more likely to produce CMEs \citep[e.g.,][]{2007ApJ...654L.171L,tor11}. Occasionally, small secondary fan-spine systems are identified in the fan structure of large fan-spine systems \citep{hou19b}, and such nested fan-spine systems were possibly apt to produce successful CMEs due to multiple null point reconnection and therefore more energy released in the magnetic system \citep{Lih19}.

In this paper, we studied two successive filament eruptions in a tripolar magnetic field region whose coronal magnetic field showed as a fan-spine magnetic system. The main aim of the present study is to figure out the eruption mechanism of the filament eruptions and the associated activities, as well as to verify whether there is the physical connection between the two successive filament eruptions. Observations and methods are introduced briefly in Section ~\ref{sec:data}; the main analysis results are presented in Section ~\ref{sec:result}; interpretation of the event is described in Section ~\ref{sec:discuss}; the conclusion and discussions are given in Section ~\ref{sec:summary}.

\section{Observations and Methods}\label{sec:data}
To explore the physic mechanism of the present event, we use high-resolution observatories taken by the Kanzelh\"{o}he Solar Observatory ({\em KSO}) and the {\em Solar Dynamics Observatory} \citep[{\em SDO};][]{pesnell12}. These observations allow us to explore the eruption details and three-dimensional (3D) coronal magnetic field environment with the aid of the Nonlinear Force-Free Field (NLFFF) model. The Atmospheric Imaging Assembly \citep[AIA;][]{lemen12} onboard the {\em SDO} observes the full-disk Sun in 10 channels with a {1.2\arcsec} spatial resolution, which included 7 extreme ultraviolet (EUV), 2 ultraviolet (UV) and 1 visible-light wavelengths. We mainly use AIA EUV images at 304 \AA\, 171 \AA\, 193 \AA\  and 94 \AA\ wavelengths with a 12 s temporal cadence, and the UV images at 1600 \AA\ with a 24 s temporal cadence. The full-disk line-of-sight (LOS) photospheric magnetograms and vector magnetic fields provided by the Helioseismic and Magnetic Imager \citep[HMI;][]{Schou2012} onboard the {\em SDO}, and their temporal resolutions are of 45 s and 720 s, respectively. KSO provides the full-disk solar images at 6562.8 \AA\ (H$\alpha$ line center) with a {2\arcsec} spatial resolution and a temporal cadence of 1 minute. The {\em Reuven Ramaty High Energy Solar Spectroscopic imager} \citep[RHESSI;][]{lin02} provides X-ray fluxes with a temporal resolution of 4 seconds. Moreover, we also use the soft X-ray (SXR) fluxes taken by the {\em Geostationary Operational Environmental Satellite} ({\em GOES}) to study the flaring process. 

\begin{figure*}[thbp]
\epsscale{0.85}
\plotone{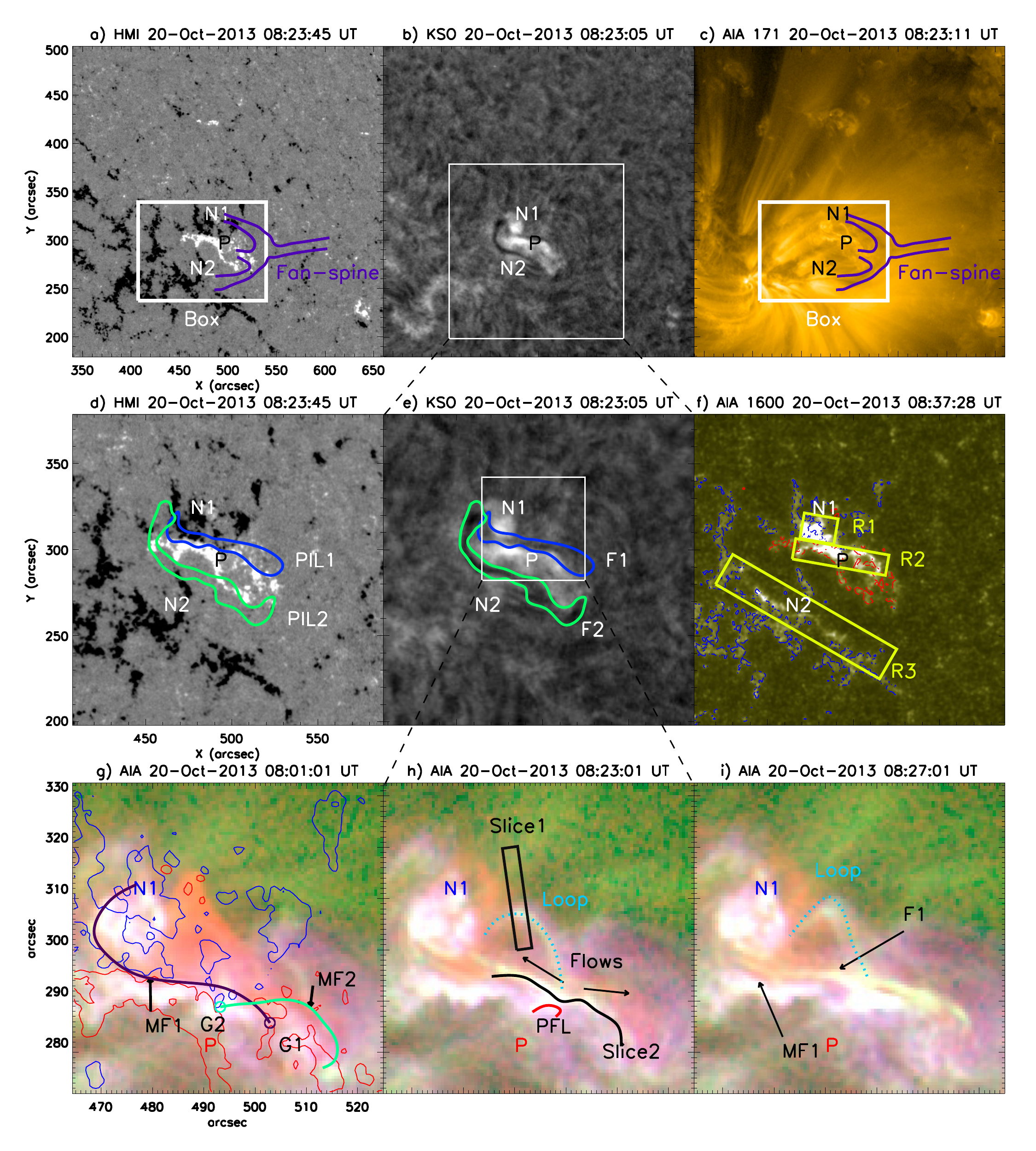}
\caption{AIA images and HMI LOS magnetogram show the signatures during the sympathetic eruptions. Panels (a)-(c): the overview of the active region, the white boxes in panel (a) and (c) are used to collect AIA light curves and magnetic fluxes in Figure~\ref{fig6}. Panels (d)-(f): the location of some features in panels (a)-(c), The blue and green curves show the approximate locations of the two filaments, and the yellow, blue and purple boxes indicate the three ribbons( R1, R2 and R3) after the eruptions,  respectively. Panel (g)-(i) are composite AIA images. Panel (g): the body of filaments are labeled by MF1 and MF2, and the LOS magnetogram at 08:23:45 UT are overlaid at $\pm100$ G on composite AIA image, where the red (blue) color indicates positive (negative) polarity. Panel (h): the features of bi-directional flows, the post flare loops (PFL), the bright loop upon the filaments. Panel (i): the arrow with F1 denotes the new-formed filament after reconnection, and the blue dotted line outlines the loops upon the filaments. An animation of the sympathetic eruption is available. The animation has a 12 second cadence,  including five EUV bands (94\AA\ 131\AA\ 30\AA\ 171\AA\ 193\AA\ ) and a UV band (1600\AA\ ) from 08:15 UT to 09:05 UT. An animation of this event is available.
\label{fig1}}
\end{figure*}

To understand the 3D coronal magnetic environment around the eruption source region, we extrapolated the 3D coronal magnetic field using the photospheric vector magnetic fields as the lower  boundary input, with the NLFFF software adopting the “weighted optimization” calculation method \citep{wht00,wig04}. The HMI vector magnetograms have a time cadence of 12 minutes, and those taken at 08:00 UT, 08:24 UT and 08:48 UT are used to extrapolate the 3D coronal magnetic fields. The eruption source region of the present event was close to the disk center; therefore, the projection effect should be weak in our extrapolation result. We select a large (grid point: {$396 \times$ 396 $\times$ 396) region to check the magnetic field connectivity of the whole eruption source region. To investigate the kinematics of various structures such as the filaments, loops and plasma ejections, we constructed a series of time-distance diagrams by using the time sequenced AIA images. To obtain a time-distance diagram, one needs to get the intensity profiles at different times along a specific path, then stacks up these one-dimensional intensity profiles in time to generate a two-dimensional time-distance diagram.
  
\section{Result}\label{sec:result}
On 2013 October 20, a {\em GOES} soft X-ray C2.9 flare took place in NOAA active region AR11868, which was accompanied by two successive filament eruptions and a CME. The flare started and peaked respectively at about 08:30 UT and 08:40 UT based on the {\em GOES} 1--8 \AA\ soft X-ray flux. The top and middle rows of \nfig{fig1} show the pre-eruption source region and the locations of the flare ribbons, while the bottom row shows the formation process of a small filament. The HMI LOS magnetograms show that the eruption source region was a tripolar magnetic region composed of a positive polarity (P) in-between two negative magnetic polarities (N1 and N2). Such a magnetic field environment forms two polarity inversion lines (PIL) where filaments are apt to form (see \nfig{fig1} (a) and (d)). By checking the KSO H$\alpha$ center images, we do find two filaments resided along the two PILs (see \nfig{fig1} (b) and (e) and the blue and green contours in \nfig{fig1} (d) and (e)). Here, for the sake of description, hereafter, we name the filament resided in-between P and N1 (N2) as F1 (F2). In EUV images, an obvious fan-spine magnetic configuration can be identified as outlined by the purple curves in \nfig{fig1} (a) and (c). An AIA 1600 \AA\ image during the eruption is displayed in \nfig{fig1} (f), which exhibits the locations of three flare ribbons that often can be expected in the eruption of fan-spine magnetic systems. The ribbons were located at the sites of the three magnetic polarities P, N1 and N2, and they are outlined by the three boxes. We did not observed the expected remote brightening in association with the outer spine, which was possibly that the remote footpoint of the outer spine was located in the back side of the solar disk, or the brightening was too weak to be detected.

In AIA observations, it is evidenced that F1 was formed by two crossed mini-filaments (MF) through magnetic reconnection between the two MFs around their crossing site. The detailed evolution process is shown by the composite tri-color images in the bottom row of \nfig{fig1}, by using the AIA 94 \AA\,171 \AA\ and 193 \AA\ images. The main axes of MF1 and MF2 are highlighted by a black and a green curves in \nfig{fig1} (g), respectively. The reconnection between MF1 and MF2 can be evidenced by several features: the sudden appearance of a small bright loop between G1 and G2 (see the red curve in \nfig{fig1} (h)), the bi-directional plasma flows from the crossing site of the two MFs (see the two black arrows in \nfig{fig1} (h)), and the formation of F1 as indicated by the black arrow in \nfig{fig1} (i). Such kind of magnetic reconnection between two filaments were observed in several previous events, which was thought to be important for the generation of two-sided-loop solar jets and the formation of filaments \citep[e.g.,][]{tian2017,2016ApJ...816...41Y,2017ApJ...840L..23X,zhengr17,yangbo19,2016ApJ...818L..27C,chenh18}. Another interesting feature during the eruption is the appearance and kinking eruption of a bright loop as indicated by the dotted blue curve (labeled with ``loop'') in \nfig{fig1} (h) and (i). The detailed eruption of this loop will be described later.  

\begin{figure*}[thbp]
\epsscale{0.85}
\plotone{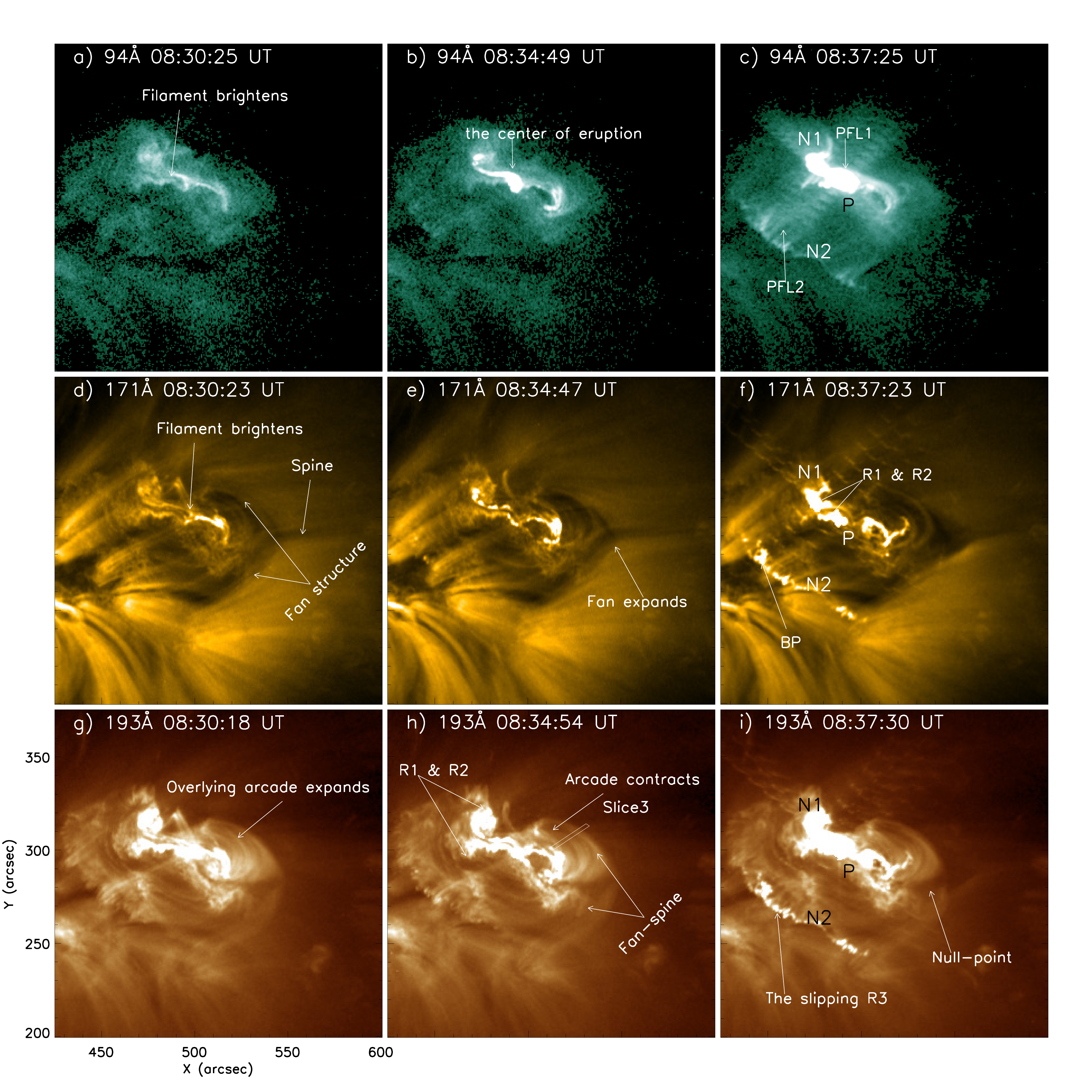}
\caption{The time sequence of the first eruption in AIA 171 \AA, 193 \AA~and 94 \AA~images, different wavebands show the dynamic evolution of the overlying loops and filament. Panels (a)-(c): AIA 94 \AA~images, panels (d)-(f): AIA 171 \AA~images, panels (g)-(i): AIA 193 \AA~images. 
The first column points out initial brightening of the F1, a well defined fan spine and the overlying loops upon F1 at 08:30 UT.
The second column denotes the center of the eruption, the parallelogram labeled with 'Slice 3' in panel (h) is used to investigating the contraction of the overlying arcades displayed in \nfig{fig6}. The third column shows the final stage of F1 eruption, including the post flare loops PFL1 (PFL2) caused by the internal (external) reconnection, the three ribbons (R1, R2 and R3). 
\label{fig2}}
\end{figure*}

The eruption of F1 and the associated eruption features are displayed in \nfig{fig2} to \ref{fig4}, as well as the animation available in the online journal. F1 started to rise at about 08:30 UT; it firstly brightened and then quickly erupted. The filament material seemingly drained back to the solar surface along its two legs, and therefore the filament body, especially the apex part, became weaker during its eruption. This filament eruption did not cause any CME signal in the outer corona, which might be collapsed or the eruption was too weak to escape the Sun. In the low corona, two groups of PFLs (PFL1 and PFL2) and three flare ribbons (R1, R2, and R3) were observed to be associated with the filament eruption. The PFLs can be best seen in the AIA 94 \AA\ images, in which PFL1 connects P and N1, while PFL2 connects P and N2 (see the top row of \nfig{fig2}). The three bright flare ribbon situated at the footpoints of the two groups of PFLs, i.e., along the magnetic polarities of P, N1 and N2. In the AIA 171 \AA\ images (see the middle row of \nfig{fig2}), the three flare ribbons can be clearly observed. The appearance of R1 and R2 were simultaneously at about 08:33 UT, while R3 appeared about two minutes later at about 08:35 UT. In addition, the coronal loops consisting the fan structure of the fan-spine system experienced an obvious expansion but only coronal loops met contraction movements during the eruption of F1. Similar process can also be identified in the AIA 193 \AA\ images (see the bottom row of \nfig{fig2}). 

\begin{figure*}[thbp]
\epsscale{0.85}
\plotone{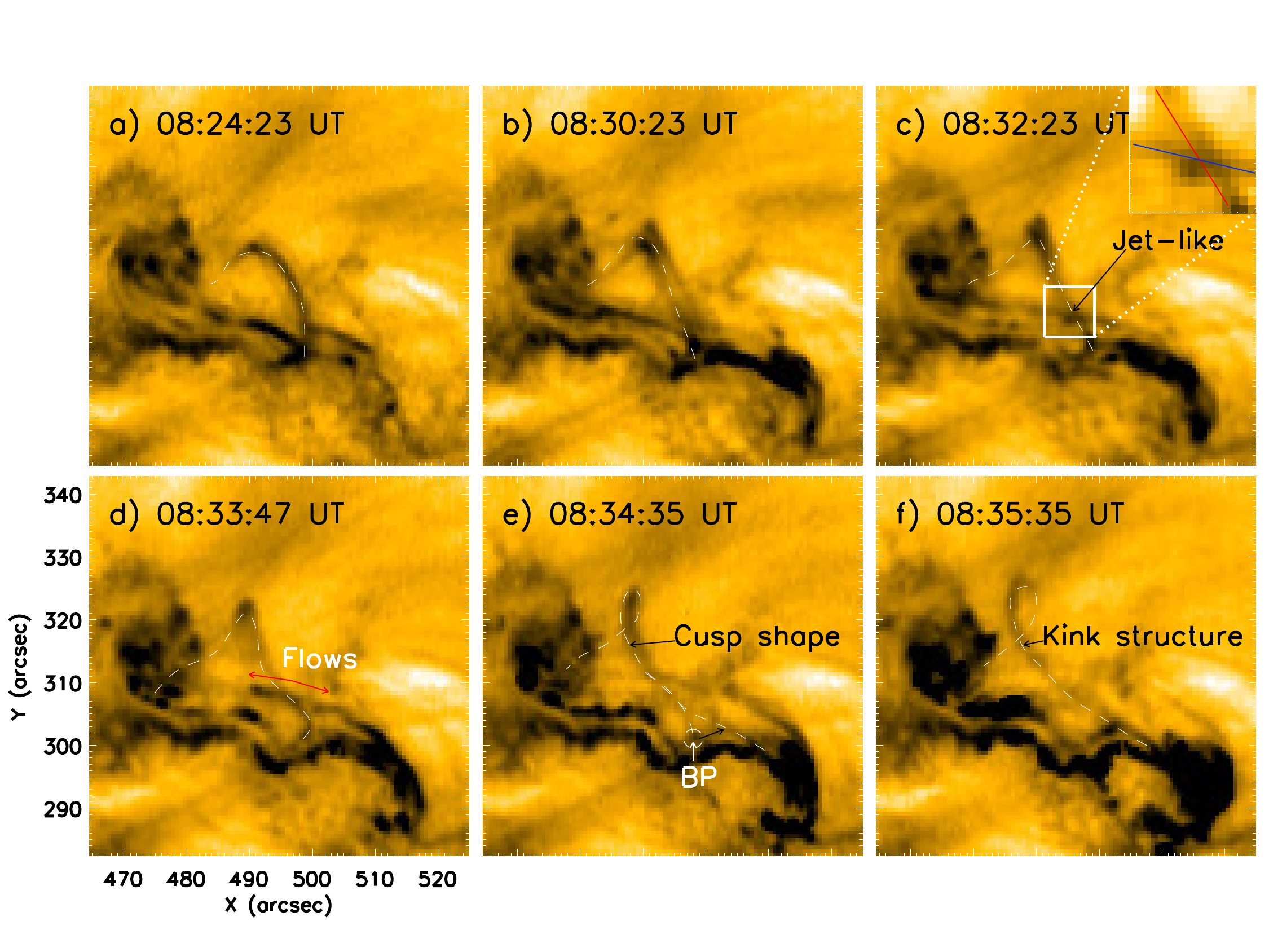}
\caption{Panels (a)-(f): the negative  AIA 171 \AA~images, while the black shows the area of strong bright emission, the dashed white lines outline the bright loop. Panel (c): the white box shows a jet-like structure, the blue and red straight lines denote the MFR of F1 and the loop. Panel (d): the bi-directional flows when the F1 crushed to the loop. Panel (e): the circular dashed line indicates the bright point during the eruption, the arrow labels its moving direction. Panel (f) : the writhed loop before eruption. An animation of the loop eruption is available. The animation has a 12 second cadence,  including one EUV bands (171 \AA\ ) from 08:24 UT to 08:37 UT. An animation of this event is available.
\label{fig3}}
\end{figure*}

\nfig{fig3} combines negative AIA 171 \AA\ images to display the eruption details of the loop labeled at \nfig{fig1}. In a negative image, the white and black features represent the dark and bright structures in the original image. At 08:24 UT, the loop showed as a normal potential shape. This figure shows some features appeared around the crossing site (see the inset and arrows in Figure 3 (c) and (e) : a jet-like structure, bidirectional flows, a bright point. A sudden change of the topology of the loop structure can also be identified there (see the animation available in the online journal). It is reasonable for the occurrence of magnetic reconnection between F1 and the loop. This loop finally violently erupted at about 08:36 UT, and it was followed by the violent eruption of F1. Due to the projection effect, here, we can not confirm whether magnetic reconnection occurred or not at the crossing point of the kink structure as what has been evidenced in a few observations \citep[e.g.,][]{2006ApJ...653..719A,shen12b,lil16}.  

\begin{figure*}[thbp]
\epsscale{0.85}
\plotone{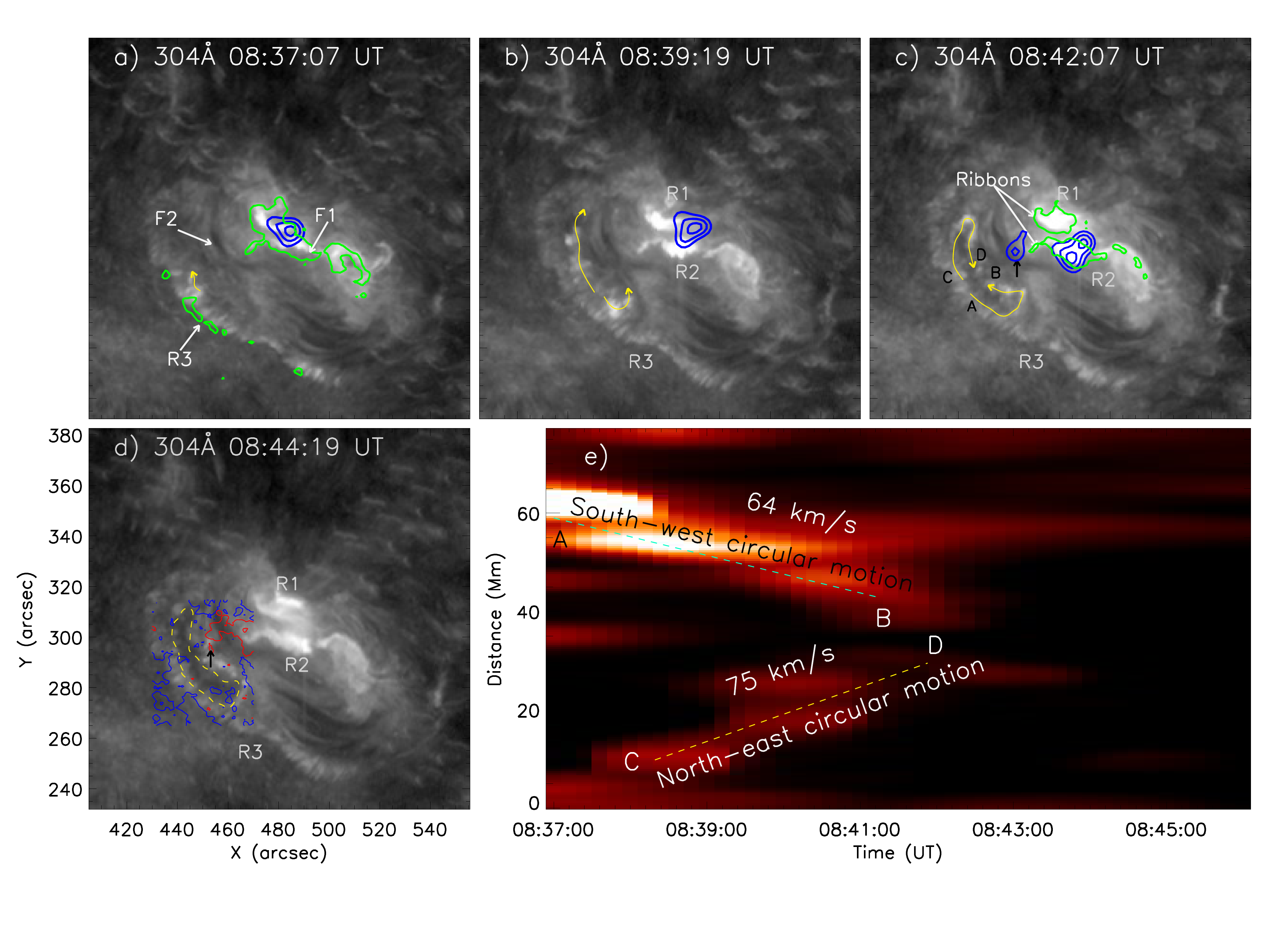}
\caption{Panels (a)–(d): Four snapshots of the AIA 304 \AA~images. In the panels (a)-(c), the yellow arrows show the 
two slipping bright ribbons at the west side of the east section of R3, the solid blue lines and the green contours symbolize the contours of the HXR emission at 12–25 keV and the intensity of 1600 \AA, respectively. The red (blue) contours in panel (d) are overlaid by the HMI magnetic field at $\pm$ 25 G. Panel (e): the time–distance diagrams obtained by the yellow dashed curve in panel (d).
\label{fig4}}
\end{figure*}

From 08:37 UT to 08:45 UT, we observed an interesting small bright circular ribbon at the west side of the east section of R3. The detailed evolution and kinematics are displayed in \nfig{fig4}, we use four snapshots of AIA 304 \AA\ images to illustrate. The blue and green contours overlaid in \nfig{fig4} (a) and (c) are obtained from {RHESSI} hard X-ray emission sources at 12--25 keV energy band and AIA 1600 \AA\, respectively. The small circular ribbon started from a sudden brightening at about 08:37 UT as indicated by the yellow arrow in \nfig{fig4} (a), then a bi-directional slipping of the brightening in northwest and southeast directions were observed along a circular path as shown by the two yellow arrows in \nfig{fig4} (b), they met at the west side of the circular path at about 08:45 UT (see \nfig{fig4} (c) and (d)).  The time-distance diagram made from the AIA 304 \AA\ images along the circular ribbon is shown in \nfig{fig4} (e), in which the slipping bright feature can be clearly identified. It is measured that the slipping speeds of the brightening in southwest and northeast directions were about \speed{64} and \speed{75}, respectively.  An additional hard X-ray source was detected at 08:42 UT, which located between the main source and the small circular ribbon (see the black arrow in \nfig{fig4} (c)). In \nfig{fig4} (d), the LOS HMI magnetogram was overlaid as contours on the AIA 304 \AA\ image. As indicated by the black arrow in \nfig{fig4} (d), there was a small positive polarity close to the north boundary of the circular ribbon. Here, we consider that the small circular ribbon and the small hard X-ray source were possibly associated with the null point reconnection of a small fan-spine system. The triggering of the small fan-spine system's eruption was probably caused by the disturbance resulted from the eruption of F1, and the slipping motion of the circular ribbon suggests the slipping magnetic reconnection in the fan quasi-separatrix layer \citep[e.g.,][]{shen19b}.

\begin{figure*}[thbp]
\epsscale{0.85}
\plotone{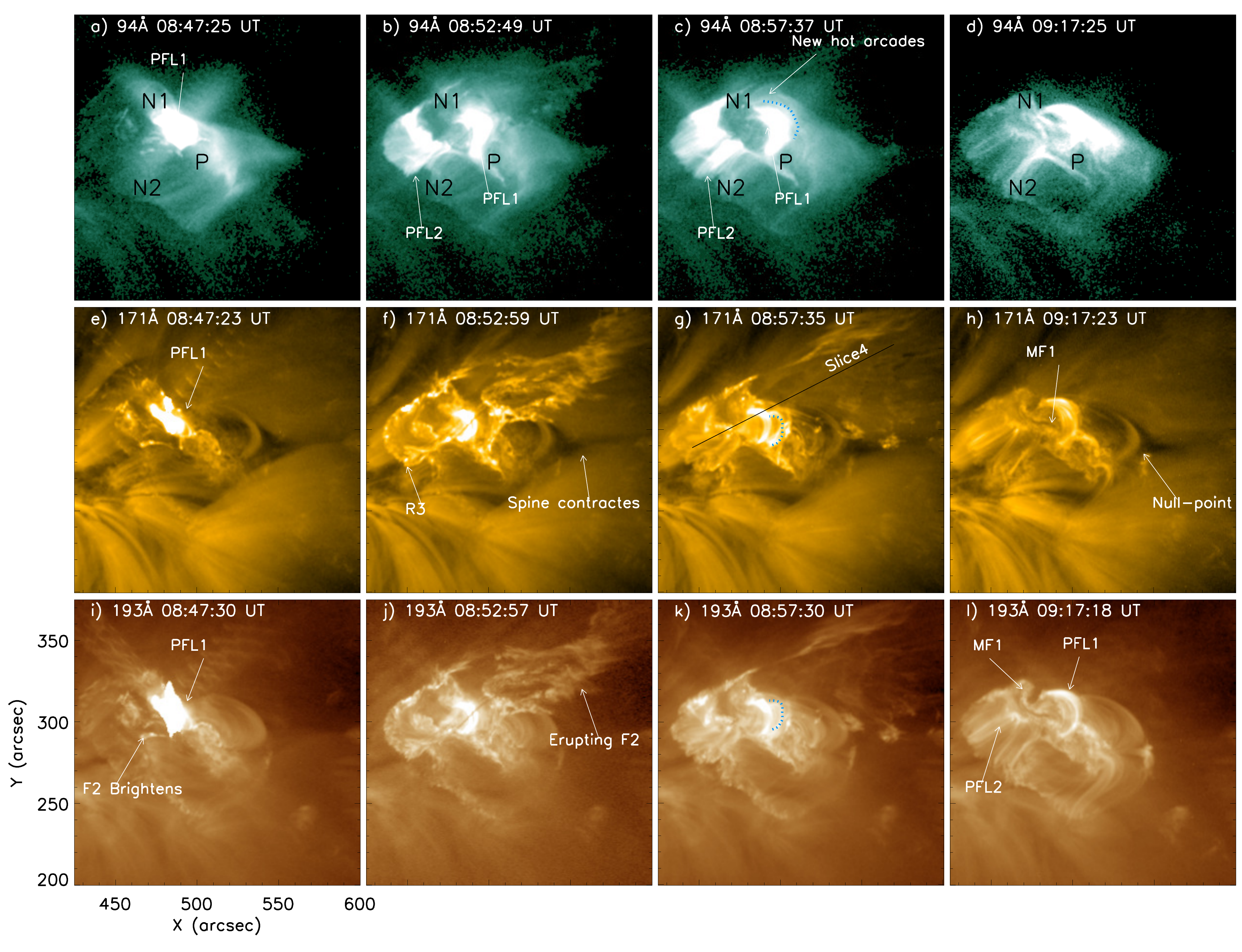}
\caption{The time sequence of the second filament eruption of AIA 171 \AA~, 193 \AA~and 94 \AA~images in log scale, panels (a)-(d): AIA 94 \AA~images, panels (e)-(h): AIA 171 \AA~images, panels (i)-(l): AIA 193 \AA~images. The arrows in the first column show the PFL1 and initial brightening at the center of the F2. The second column displays the eruptive phase of F2, the arrows mark the features of the PFL1, PFL2, R3 and the contraction of spine. The blue dashed lines in the third column point to the new hot arcades during the eruption. The solid black line with 'Slice 4' in panel (j) is used to investigating the projection velocity of the CME. The fourth column shows the final stage of the eruption, the arrows point out the MF1, PFLs (PFL1 and PFl2) and the null point of the fan-spine.
\label{fig5}}
\end{figure*}

Right after the total formation of the small circular ribbon, F2 started to rise at about 08:45 UT and then erupted violently at about 08:48 UT. The eruption process is displayed in \nfig{fig5} with the AIA 94 \AA\, 171 \AA\ and 193 \AA\ time sequence images. The eruption of F2 underwent a slow rising phase during 08:45 UT to 08:48 UT, then it erupted quickly to the northwest direction and caused a CME in the outer corona (the CME is not displayed). At about 08:48 UT, PFL1 associated with the eruption of F1 was still very bright, but PFL2 had disappeared. At the beginning of F2's violent eruption phase, brightening firstly appeared below the filament, then a large bright loop system connecting opposite magnetic polarities of P and N2 appeared (PFL2). It is clear that this hot loop system was the PFL caused by the eruption of F2, due to the magnetic reconnection between the two legs of F2's confining magnetic field lines below the filament. It is interesting that new hot loops were continuously formed at the site of PFL1, connecting magnetic polarities of P and N1. By considering the fan-spine magnetic configuration of the eruption source region, these newly formed hot loops should be the consequence of magnetic reconnection around the null point of the main fan-spine system, between the confining fields of F2 and the magnetic reconnection favorable open field lines around the null point. During the eruption of F2, the fan structure also showed a first expansion and then a contraction motion. Although this filament eruption caused a CME in the outer corona, the fan-spine system was not completely destroyed. On the contrary, it recovered its initial configuration before the eruptions (see \nfig{fig5} (i) and (p)).

\begin{figure*}[thbp]
\epsscale{0.85}
\plotone{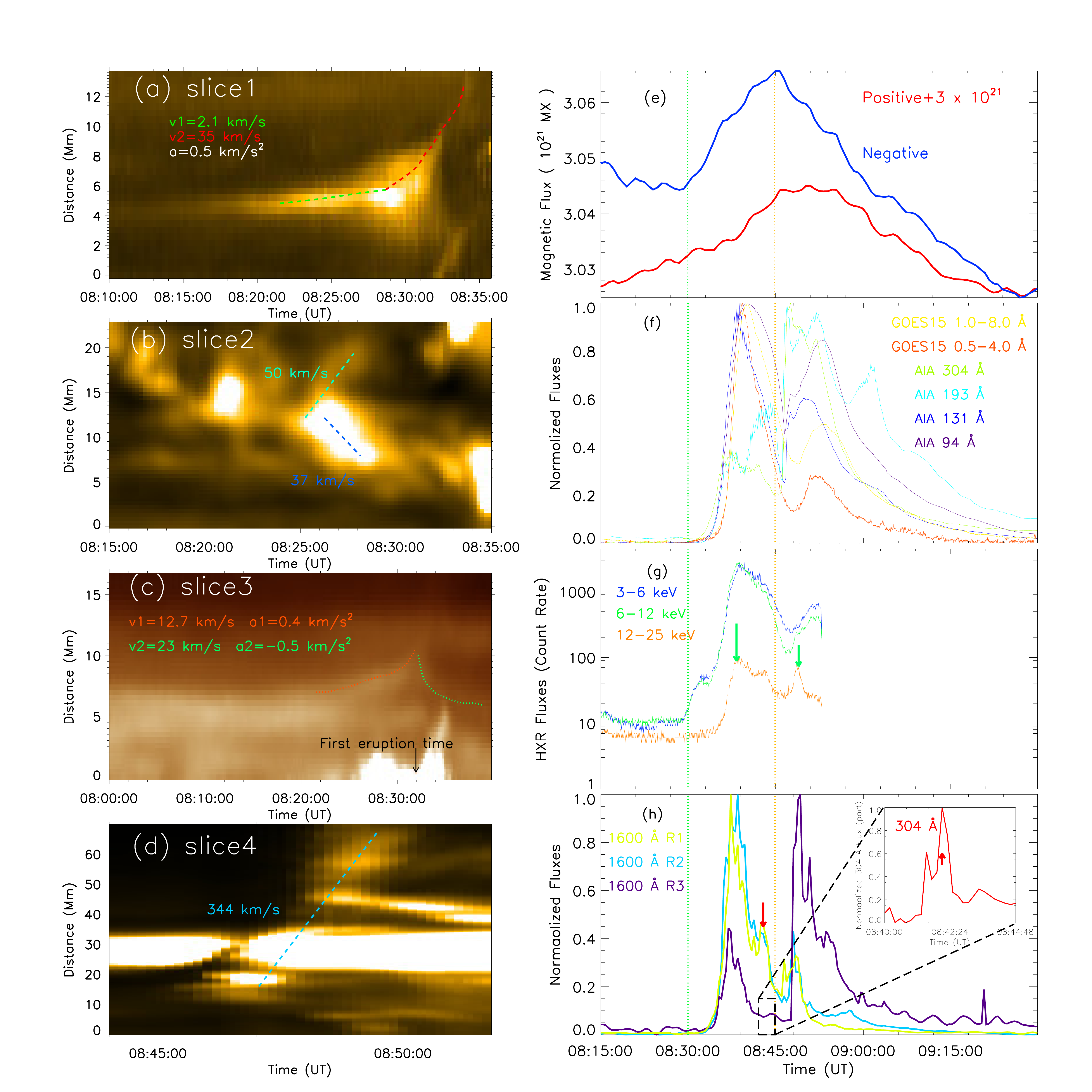}
\caption{The left four panels (a)-(d): time–distance diagrams obtained by slice1 and slice2 in Figure~\ref{fig1}, slice3 in Figure~\ref{fig2} and slice4 in Figure~\ref{fig5}. Panel (a): projection velocity of the eruptive loop. 
Panel (b): made from AIA 171 \AA~, the solid blue and green lines point to the velocity of the bi-directional flows. 
Panel (c): the red and green curves indicate the trajectory of the overlying arcades of the F1, the black arrow labels the first eruption time.
Panel (d): made from AIA 171 \AA~, the purple lines stand for the rising velocity of the F2 caused by the second flare.
The right four panels (e)-(g): SXR, HXR, AIA light curves of the two sympathetic filaments eruptions and temporal evolutions of magnetic flux with the box in panel (a) of Figure~\ref{fig1}, panel (h) shows the light curves of ribbons (R1, R2 and R3) associated with the filaments eruption.
\label{fig6}}
\end{figure*}

The time-distance diagrams along slice1 to slice4 are displayed in the left column of \nfig{fig6} to study the dynamic evolution of the event. \nfig{fig6} (a) is the time-distance diagram along slice1 (shown in \nfig{fig1} (c2)), which is used to study the eruption of the kink loop stretching over F1. The loop experienced a slow rising phase before its violent eruption at about 08:29 UT. The mean speeds during the slow rising and violent eruption phases were about \speed{2.1 and 35}, respectively. During the transition phase from the slow rising to the violent eruption phase, it had an acceleration of about \accel{0.5}. The time-distance diagram along slice2 (shown in \nfig{fig1} (c2)) is shown in  \nfig{fig6} (b), which shows the bi-directional ejections of plasma flows caused by the magnetic reconnection between MF1 and MF2. It can be seen that the plasma ejections started at about 08:25 UT, and the ejecting plasma flows in the east and west directions were about \speed{50 and 37}, respectively. The expansion and contraction motions of the north lobe of the main fan-spine are examined by using a time-distance diagram along slice3 (see \nfig{fig2} (h)), and the result is displayed in \nfig{fig6} (c). The time-distance diagram clearly shows the expansion and contraction motion of the loop system. It is noted that the transition from expansion to contraction motion occurred at about 08:32 UT, 2 minutes after the start of F1's slow rising. The speed and acceleration of the expansion motion were about \speed{12.7} and \accel{0.4}, while the contraction motion were about \speed{23} and \accel{0.5}, respectively. \nfig{fig6} (d) shows the time-distance diagram along slice4 (see \nfig{fig5} (j)), from which we obtain the eruption speed of F2 was about \speed{344}, consistent with the average speed  (\speed{488}) of the associated CME whose first appearance in the FOV of LASCO C2 was 09:12 UT \footnote{\url{http://sidc.oma.be/cactus/catalog/LASCO/2_5_0/2013/10/CME0092/CME.html}}.

The right column of \nfig{fig6} shows the plots of various fluxes in association with the event, in which the two vertical dotted lines indicate the start times of the two filament eruptions at 08:30 UT and 08:45 UT, respectively. \nfig{fig6} (e) shows the variations of positive (red) and negative (blue, absolute value) magnetic fluxes within the white box as shown in \nfig{fig1} (a). The positive magnetic flux showed an increasing trend until at about 08:45 UT, then it kept a constant value for about 10 minutes before the start of rapid decreasing at about 08:55 UT. For the negative magnetic flux, it showed an increasing trend from 08:30 UT to 08:45 UT, then it changed to a rapid decreasing trend until the end of the event. The soft X-ray {\em GOES} fluxes in the energy bands of 0.5--4 \AA\ and 1--8 \AA\ , EUV intensity fluxes of AIA 304 \AA\, 193 \AA\, 131 \AA\ and 94 \AA\ within the eruption source region (see the white box in \nfig{fig1} (c)). The {\em RHESSI} X-ray fluxes in the energy bands of 3--6 keV, 6--12 keV, and 12--25 keV. \nfig{fig6} (h) shows the intensity fluxes within the three boxes in \nfig{fig1} (f), which reflect the temporal intensity variation of the three flare ribbons. Based on these flux curves in \nfig{fig6} (f)--(h), one can identify two main peaks at about 08:38 UT and 08:49 UT. The two main peaks well reflected the eruption of the two filaments, and their start times were at about 08:30 UT and 08:45 UT, respectively. In addition, a small peak can also be identified at about 08:42 UT (see the red arrow and the inset in \nfig{fig6} (h)), three minutes before the onset of the second filament eruption. Based on the imaging observational results, this small peak can be recognized as the evidence of the eruption of the small circular ribbon as described in \nfig{fig4}.

\begin{figure*}[thbp]
\epsscale{0.85}
\plotone{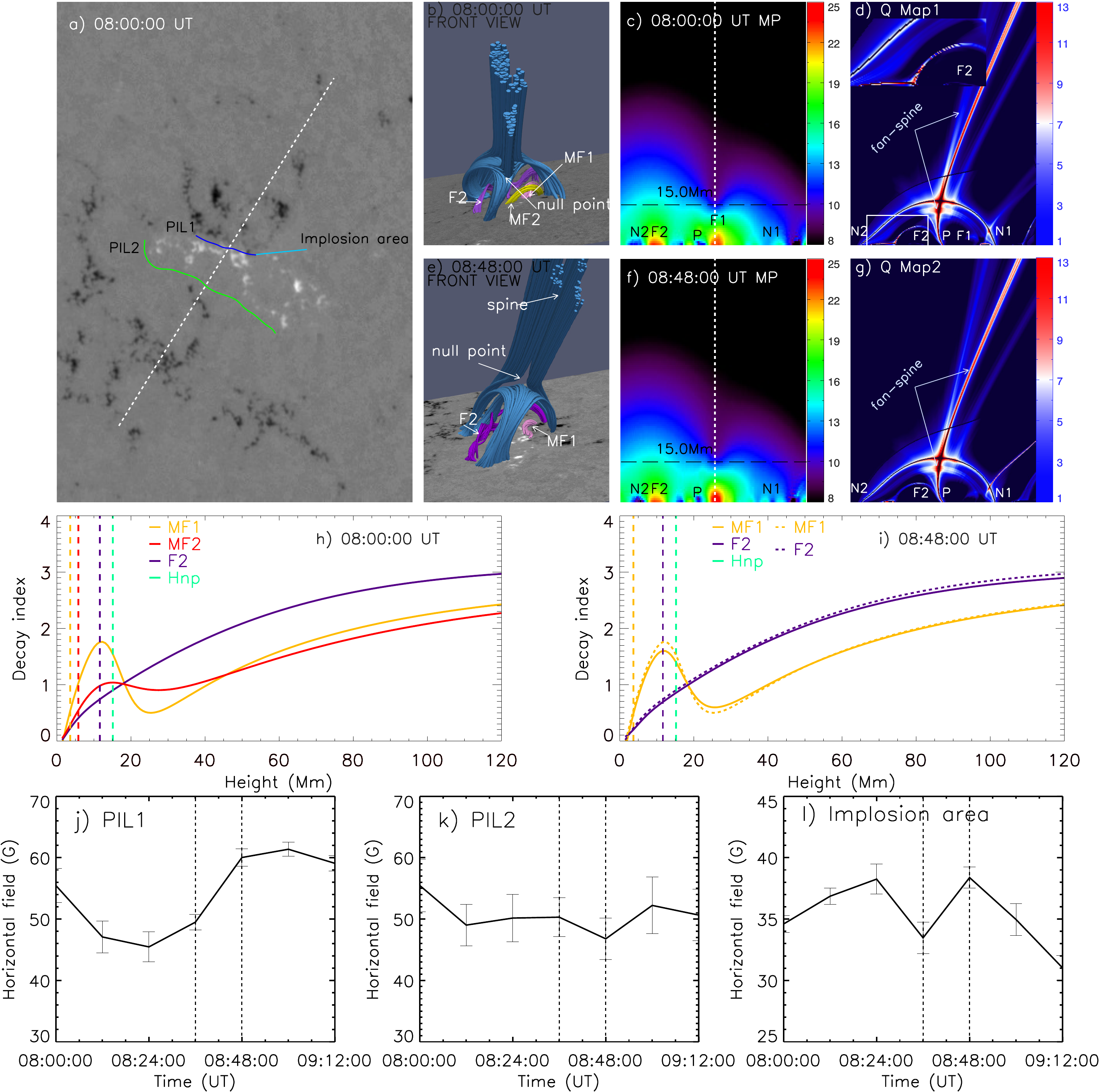}
\caption{Panel (a) displays a HMI LOS magnetogram at 08:00:00 UT, while the dashed line indicates the location of the cross-section where we calculate the map of magnetic pressure (panels (c) and (f)) and squashing factor Q (panels (d) and (g)), and the colored lines display the location of the PILs and the site of the implosion. Panels (b) and (e) show some overlying arcades of the fan-spine structure (blue) and the filaments (MF1 (pink), MF2 (yellow) and F2 (purple)) at two pre-eruptive phases from the front view. The black dashed lines and white dotted lines in panels (c) and (f) display the location of null point. The arrows in panels (d) and (g) indicate the fan-spine structure. Panels (h) and (i) illustrate the decay index above the filaments, the vertical dotted lines point to the heights of MF1 (yellow), MF2 (red), F2 (purple) and null point (green), respectively. Panels (j)-(l) display the horizontal fields across the PILs and explosion area in corona at the height of 7 Mm to 15 Mm, the standard errors for each measured lines are about $\pm$1.8 G, $\pm$4 G, $\pm$1 G, and the black dotted lines in the three panels are the eruption time of the two filaments.
\label{fig7}}
\end{figure*}

Using the HMI photospheric vector magnetic fields as input, the 3D coronal magnetic fields were extrapolated to study the magnetic structure around the eruption source region. Since the NLFFF model is not a good way to obtain the coronal magnetic field of active regions during the impulsive phase of flares, we chose the vector magnetic fields at 08:00 UT, 08:24 UT (before the event) and 08:48 UT (before the eruption of F2) to extrapolate the 3D magnetic fields. The extrapolated results are shown in \nfig{fig7} and \nfig{fig8}. \nfig{fig7} (a) shows the HMI LOS magnetogram around the eruption source region, in which the black dotted line shows the position where we calculate the magnetic pressure and squashing factor maps in the height direction. Panels (b)--(d) show the magnetic field lines, magnetic pressure, and Q-map calculated based on the NLFFF extrapolation result at 08:00 UT, panels (e)--(g) are the corresponding maps calculated based on the extrapolated magnetic field at 08:48 UT, and panels (j)-(k) are the horizontal fields in corona calculated by NLFFF models across the colored lines in panel (a). The first two columns of \nfig{fig8} show the flux ropes and twist maps corresponding to filaments, beside, the predicted ribbons according to the squashing factor Q on the surface are also displayed in the third column. Panels (j)--(k) are the photospheric transverse filed cross the PFLs (PFL1 and PFL2). The extrapolated coronal magnetic field at 08:00 UT well reveals the fan-spine magnetic system and the filaments as those observed in the direct imaging observations, i.e., F1 (form by the coalescence of MF1 and MF2) and F2 located in the north and south lobes of the inner fan structure, respectively (see \nfig{fig8} (a)).  At 08:24 UT, from the 3D magnetic configuration, the F1, MF1 and loop appear. The loop is originated in the south positive polarity and mostly connected upon the filaments, indicated by the green lines in the \nfig{fig8} (d), whose profile in the SDO views are very similar to the shape of the loop in \nfig{fig3} (a). At 08:48 UT, F2 and MF1 were also revealed by the extrapolated magnetic field, but MF2 can not (see \nfig{fig8} (g)). Compared to the magnetic field lines in 08:00 UT and 08:48 UT, The left lobe of fan-spine gets its inner spine closer to the right spine. The extrapolated results are in agreement with the direct imaging observational results as shown in the bottom of \nfig{fig1} and \nfig{fig2} (e). 

\nfig{fig7} (c) and (f) show the magnetic pressure maps at 08:00 UT and 08:48 UT along the black dotted line (see \nfig{fig7} (a)) in the height direction, respectively. The locations of F1 (or MF1 and MF2) and F2 can be identified as high magnetic pressure regions (red color) well below the north and south lobes of the fan structure, respectively. The null point of the fan-spine magnetic system shows as a small region of low magnetic pressure, and the height of its center is about 15 Mm above the solar surface (see the horizontal dashed line in \nfig{fig7} (c) and (f)). The maps of squashing factor Q in the height direction along the black dotted line (see \nfig{fig7} (a)) at 08:00 UT and 08:48 UT are displayed in \nfig{fig7} (d) and (g), respectively. Here, we use the code developed by \cite{liu2016} to generate the Q-maps. The Q-maps well exhibit the skeleton of the fan-spine system, in which the linear high Q regions represent the quasi-separatrix layers within the fan-spine magnetic system. It is interesting that the Q-map at 08:00 UT also revealed a small fan-spine magnetic system hosted within the south lobe of the inner fan of the main fan-spine magnetic system (see the white box and the inset in \nfig{fig7} (d)), which also indicated that F2 was below the outer spine of this small fan-spine magnetic system. The footpoint of the fan of this small fan-spine structure well corresponds to the location of the observed small circular ribbon as described in \nfig{fig4}. Therefore, the appearance of the small circular ribbon evidenced the eruption of this small fan-spine structure, in which magnetic reconnection occurred around the null point. Since F2 located below the outer spine, the eruption of this small fan-spine structure would lead to the decrease of the confinement capacity of the overlying magnetic field of F2. This could be the reason why the slow rising of F2 occurred right after the formation of the small circular ribbon.  The horizontal fields upon PILs and implosion area are measured the average value with the height from 7 Mm to 15 Mm,  which is below the null point. We estimate the measuring errors to be related to the adjacent eight pixels of the points where we choose to calculate the coronal horizontal field, and the standard errors for the calculation regions (PIL1, PIL2 and implosion area) are about 1.8 G, 4 G, 1 G, respectively. We label the start time of two filament eruptions by black dashed lines, in under the two sympathetic eruptions interval, the magnetic pressure upon the PIL1 shows a gradual increasing from 49 G to 60G, while PIL2 experiences a sightly decreasing. And the drift of transverse magnetic field upon the implosion site down from 38 G towards 33 G at the pre-eruptive phase of F1, then it is back to 38 G, the disparity of the value may be the reason for the loops contraction in \nfig{fig2}. It should be noted that the increasing horizontal field is often related to the impulsive phase of the filament eruption\citep[e.g.,][]{2019ApJS..240...11P,2012ApJ...745L..17W}, as an associated phenomenon with the eruptions. Thus, before 08:48 UT, the small fan-spine structure had just erupted, the extraordinary decreasing upon PIL2 might be caused by the magnetic topology change. We also compute the QSLs from squashing factor Q on the surface to predict the flare ribbons in both the potential and NLFFF models. The results are displayed in \nfig{fig8} (c), (f)  and (i), where the green (blue) contours are from the value of NLFFF (potential) model. One can see that both the NLFFF QSLs and the potential field QSLs are in well accordance with the brightening in 304 \AA\ images, while the potential field QSLs are more associated with the ribbons (R1, R2 and R3).

Based on the extrapolated 3D magnetic fields, the decay index of the external potential magnetic field over F1 (or MF1 and MF2) and F2 are calculated using the formula $n = -dln|B|/dln|z|$ \citep{kliem06}. Here, $B$ and $z$ are the external potential magnetic field strength and height above the solar surface, respectively. The calculated decay indexes are plotted in \nfig{fig7} (h) and (i), which are the results calculated from the extrapolated coronal magnetic fields at 08:00 UT and 08:48 UT, respectively. The vertical yellow, red, purple, and green dashed lines indicate respectively the heights of MF1, MF2, F2, and the fan-spine null point above the solar surface, while the decay indexes for MF1, MF2, and F2 are plotted as yellow, red, and purple curves. At 08:00 UT, the decays index for MF1 (MF2) shows a quick increase to about 1.7 (1.1) below the null point, and then it rapidly decreases to about 0.5 (0.9) at a height of about 25 Mm. Above the height of about 30 Mm, the decay indexes above MF1 and MF2 gradually increase to about 2.4 and 2.2 at the height of 120 Mm. Such a decay index variation pattern suggests the complexity of the overlying coronal magnetic field above MF1 and MF2 (i.e., F1), the saddle-like shape of the decay index curves mainly caused by the null point distribution overlying the filaments. The decay index above F2 shows a simple increasing trend as the increase of the height, which indicates the normal potential magnetic field distribution above F2. These decay index curves reflect the magnetic field distribution as what have been revealed by the extrapolated magnetic field lines, magnetic pressure and the squashing factor Q maps. According to the previous statistical and theoretical investigation, the critical value of the decay index for a successful (failed) filament eruption is generally greater (smaller) than 1.5 at a height of about 42 Mm from the solar surface \citep[e.g.,][]{Tor05,kliem06,Liuy08,Liu12}. For the present event, one can see that all the decay indexes above 42 Mm increase smoothly as the increasing height, and the decay index above F2 always greater than F1 (i.e., MF1 and MF2) at any specific height. At the height of 42 Mm above the solar surface, the decay index values for F1 and F2 are about 1 and 2, respectively. This result suggests that F1 (F2) was stable (unstable) for torus instability, in agreement with our observation. The decay indexes of above MF1 and F2 at 08:00 UT and 08:48 UT are compared in \nfig{fig7} (i). At these two moments, the decay index above MF1 show little difference, but that for F2 was decreased at 08:48 UT. 

\begin{figure*}[thbp]
\epsscale{0.85}
\plotone{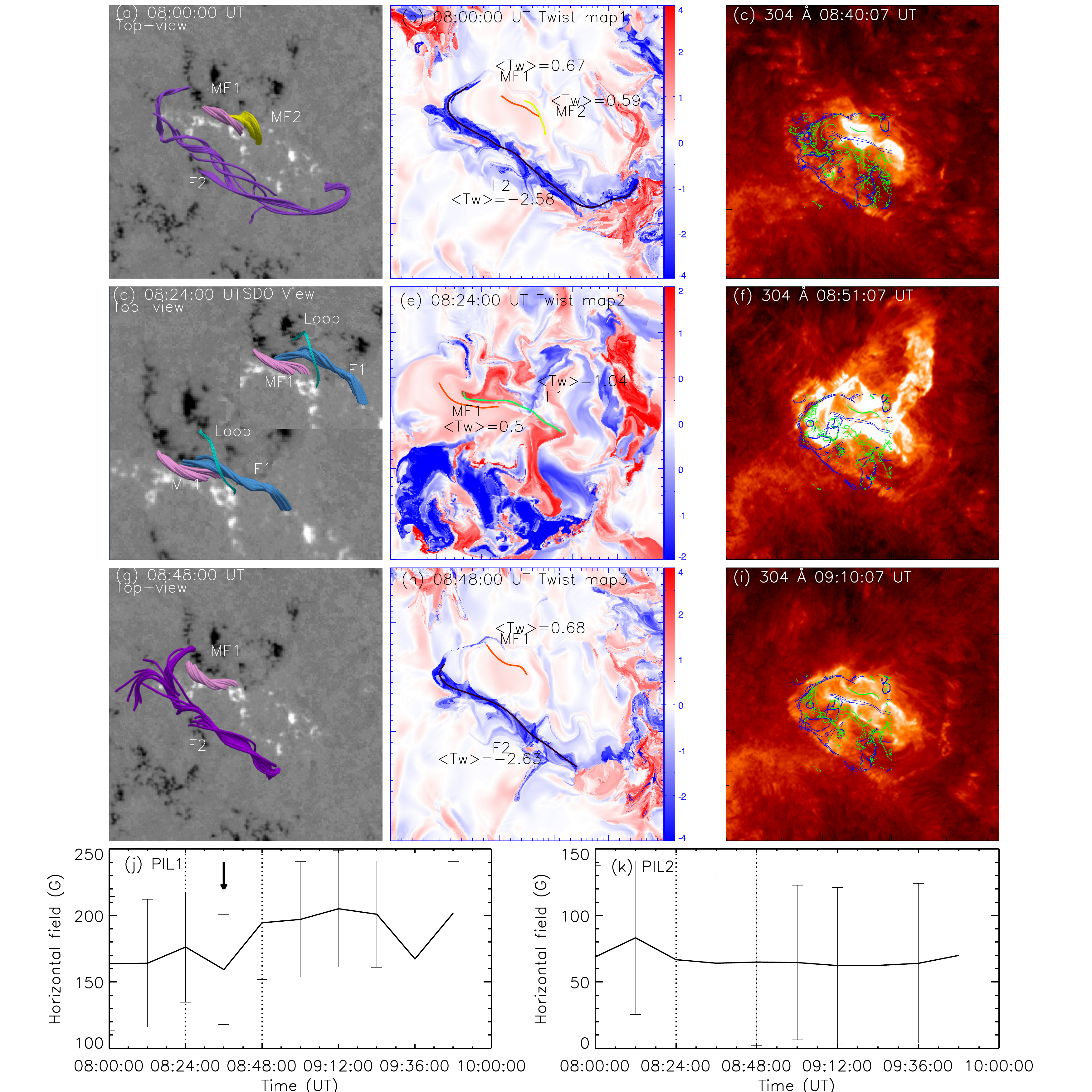}
\caption{Panel (a), (d) and (g): the top-view of the filaments from NLFFF calculation, the field-of-view (FOV) for panel (a) and (g) is 120\arcsec$\times$120\arcsec, and the FOV for panel (d) is 75\arcsec$\times$75\arcsec. Panel (b), (e) and (h): twist maps correspond to the FOV of the first column. Panel (c), (f) and (i): the QSLs from Logarithmic Q, the blue contour is calculated from the potential model, while the green is from NLFFF model. Panel (j)-(k): The photospheric transverse field across PILs (PIL1 and PIL2) as the function of time, the two black dotted lines in panel (j)-(k) label the activated time of the two filaments, and the implosion time is marked by the black arrow in panel (j), the mean uncertainty estimate for PIL1 is about $\pm$42 G, while the mean uncertainty estimate for PIL2 is about $\pm$62 G.
\label{fig8}}
\end{figure*}

In order to make sense the the mechanisms to destabilize F1 and F2, we performed the twist maps of the filaments as well as the photospheric transverse field across the PILs (PIL1 and PIL2) during the eruptions, the results are displayed in \nfig{fig8}. We also plot the location of the filaments in twist maps by colored lines and record the mean $T_{\omega}$ cross those lines. For F1, we notice that the increasing magnetic twist across the newly formed F1 (1.04) compares to the previous filaments MF1 (0.67) and MF2 (0.59). For F2, it becomes more winding and extends another branch on the body after F1 is activated. One can see that both in the two pre-eruptive phase, the MFR of F2 is featured by the dark blue color. As the eruption process, the magnetic twists of F2 gradually increases and reaches to -2.63 while the F1 maintains a low level, indicating that the eruption of F2 may be caused by torus instability and the F1 may not. From the calculation of photospheric transverse field  across the PILs in panels (j) and (k), the two black dashed lines show the activated time of the filaments F1 and F2, the sympathetic eruptions pushed a prominent $B_t$ drifts up on PIL1, marked by black arrows. The magnetic implosion started at the time around 08:35 UT, before the eruptive phase of F2, transverse magnetic field of the PIL1 drifts from 159 G towards 195 G, increases about 30\%. From the physical interpretation of the implosion according to \cite{hu00}, due to conservation of the momentum, the photospheric transverse field $B_t$ at implosion site would increase first, and then the horizontal field in corona would decrease after the implosion occurred. In \nfig{fig7} (l), one can see that the transverse magnetic field upon the implosion region has decreased at 08:36 UT, this scenario may indicate that the implosion was contemporaneous with the first eruption. We believe that gravitational and kinetic energy of the filament eruption was came from the magnetic energy releasing which would lead to the decreasing of magnetic pressure and performance for  a reduction of the transverse magnetic field in corona. Together those results provide an important evidence that with the eruption of F1 going on, the overlying loops upon F2 could move into the reconnection favor part of the right lobe (upon F1) of the fan-spine. Thus the magnetic topology change for the right lobe of fan-spine system might play an important role in the instability of F2.

\section{Interpretation}\label{sec:discuss}
Based on our analysis results, it is clear that the two successive filament eruptions originated from different locations were physically connected to each other within a tripolar magnetic field region that has a fan-spine coronal magnetic topology. We propose that the physical linkage between the two filament eruptions could be the magnetic topology change. 

\begin{figure*}[thbp]
\epsscale{0.85}
\plotone{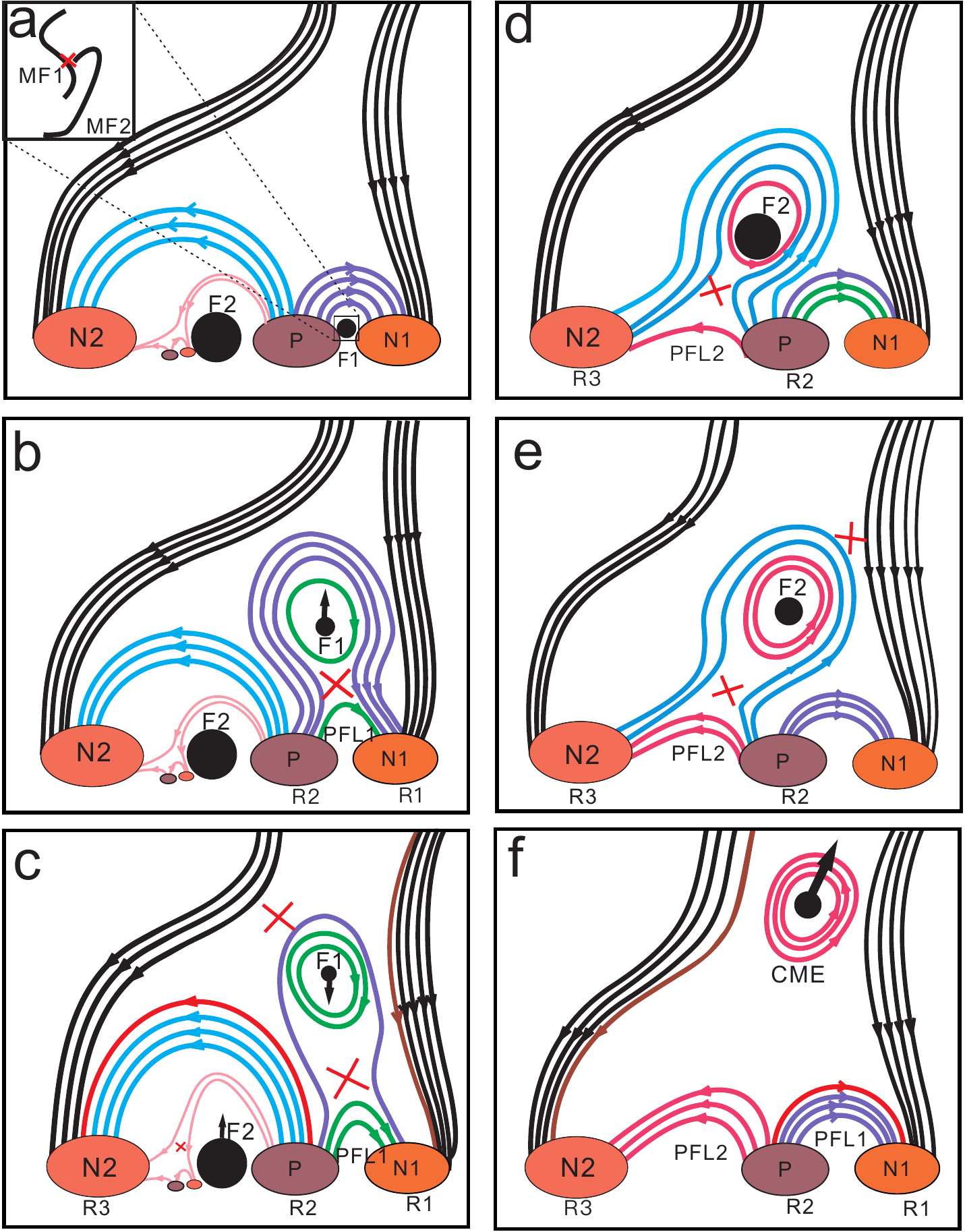}
\caption{Cartoons show our understanding of how is the sympathetic eruptions triggered. Panel (a) shows the initial magnetic configuration, the image set at the top left corner shows the original of the F1. Panel (b): the rise of F1 and internal reconnection. Panel (c): the external reconnection between the confined loops with the outer spine, the reconnection of the small fan-spine structure, and F2 begins to rise. Panel (d) illustrates the internal reconnection between the two legs of confined loops upon F2. Panel (e):  external reconnection of the confined loops upon F2. Panel (f) shows the ultimate stage of the fan-spine configuration in Figure~\ref{fig5}, the large scale CME triggered by the eruption of F2.
\label{fig9}}
\end{figure*}

We draw a cartoon in \nfig{fig9} to explain the detailed eruption and evolution processes of the sympathetic filament eruptions, in which only some representative magnetic field lines are plotted, and features including the filaments and magnetic polarities are the same as those revealed by the direct imaging observations and the extrapolated 3D coronal magnetic fields. \nfig{fig9} (a) shows the initial status of the pre-eruption magnetic configuration, i.e., a fan-spine magnetic system with two filaments reside in the two lobes, respectively. Here, we only show the details of the fan structure of the entire fan-spine system to study the physical linkage between the two filament eruptions. It should be noted that the filament (F1) between P and N1 are actually the two crossing mini-filaments as what is shown in the inset at the upper left corner of \nfig{fig9} (a). As has been shown in the bottom row of \nfig{fig1}, F1 was formed by the coalesce of MF1 and MF2 through magnetic reconnection around the crossing site. In addition, the small fan-spine structure below the left lobe of the main fan-spine system is plotted as pink curves, and with F2 being confined by its outer spine. The kinked loop structure presented in \nfig{fig3} is not plotted in the cartoon.

The newly formed F1 was unstable, and the eruption of the kinked loop as analyzed in \nfig{fig3} can further contribute to the unsteadiness of F1. According to the calculated decay index and transverse field above MF1, the unstable of F1 was probably due to the reconnection of MFs. The rising of F1 leads to a current sheet below F1 between the two legs of the confining field lines, and the magnetic reconnection (internal) in this current sheet will accelerate the eruption of the filament, result in the formation of two conjugated flare ribbons at P (R2) and N1 (R1), and the hot PFL1 connecting P and N1 (see \nfig{fig9} (b)). Due to the continuous rising of F1, the confining magnetic field lines above F1 will inevitably interact with the open field lines rooted in N2. This can lead to the formation of a new current sheet between the closed confining field lines and the open field lines rooted in N2. Consequently, the magnetic reconnection (external) in this new current sheet further removes the confining magnetic field lines of F1 and accelerates the eruption of the filament. Another consequence of this external magnetic reconnection is the formation of PFL2 connecting P and N2 and the flare ribbon R3 at N2 (see \nfig{fig9} (c)). This process can reasonably explain the delayed appearance of PFL2 (R3) for about 2 minutes with respect to the first appearance of PFL1 (R1 and R2). According to our observations, the eruption of F1 was a failed filament eruption that did not result in any detectable CME in the outer corona, which could possibly attribute to the low decay index of the confining magnetic field over F1 (see \nfig{fig9} (h), i.e., $n ~\textless ~1.5$ at the height of about 20 -- 60 Mm above the solar surface ), and the low magnetic twist as well as the strong background magnetic field (see \nfig{fig8} (e) and (k)).

How does the eruption of F1 lead to the eruption of F2? Here, we propose that this was due to magnetic topology change above F2 (likely the fan structure), causing the torus instability of F2 , and to set off the much larger eruption. For the present case, all the eruption of F1, the internal and the external magnetic reconnections as indicated by the red cross symbols in \nfig{fig9} (c), which can be explained by the three ribbons caused by standard slipping or slip-running reconnection. Then the confining loops of F2 will become unstable due to topology change caused by the F1 eruption (see \nfig{fig8} (k)). Topology change affects the stability of small fan-spine structure whose outer spine acting as a part of the confining magnetic field of F2 (see \nfig{fig9} (c)), then the eruption of this small fan-spine structure will further destroy the balance of forces acting on F2. Therefore, F2 starts to rise and triggers the internal magnetic reconnection below F2 between the two legs of the confining magnetic field lines (see \nfig{fig9} (d)). It should be noted that as the low magnetic twist and decay index, we deduce that the eruption of F1 may be a failed kink eruption, it does not deplete the material and fields, and most of its mass may be fallen down to the surface of solar, so we remain some confined loops and material in \nfig{fig9} (c). The internal magnetic reconnection will accelerate the rising of F2, causing two conjugated flare ribbons at P and N2, and a group of hot PFL connecting P and N2. The continuous rising of F2 will result in the formation of a new current sheet between the confining fields of F2 and the open field lines rooted in N1, and the magnetic reconnection (external) in this current sheet further accelerates the eruption of F2. This external magnetic reconnection also causes two conjugated flare ribbons at P and N1, and a group of hot PFL connecting P and N1 (see \nfig{fig9} (e)). Finally, the complete eruption of F2 causes a broad CME in the outer corona as what we have evidence in the coronagraph white-light images (see \nfig{fig9} (f)). The decay index above F2 was greater than the critical value of 1.5 for torus instability; therefore, the eruption of F2 was liable to become a successful eruption. It's noted that there also exists alternative possibility. According to Hudson (2000), the magnetic pressure around these energy releasing sites should be decreased during the initial stage of the magnetic reconnections, and such a process can typically last for several minutes, which are believed to be an available way to change the magnetic topology~\citep[e.g.,][]{lw2010,gos12,Sun12,sim13,rus15,wang16}. \cite{shen12a} firstly proposed that the magnetic implosion can change the magnetic topology between sympathetic filament eruptions within the framework of quadripolar breakout magnetic system. From the observation of the detected contraction of the confining loops (see \nfig{fig6} (c)), we conclude that the magnetic implosion were possibly took place upon F1. Therefore, although there is no enough evidence to demonstrate quantitatively how implosion of F1 would influence the eruption, we presume that the magnetic topology change may be the different variation in magnetic pressure caused by the magnetic implosion, the implosion results in the upper parts (stable magnetic pressure) upon F2 moving toward the implosion site (decreasing magnetic pressure) to keep the equilibrium of the system. Consequently, this can also lead to the instability and eruption of the small fan-spine structure. 
 
\section{Conclusions and Discussions}\label{sec:summary} 
Using high spatiotemporal multi-wavelength observations taken by space-borne and ground-based telescopes, we studied the sympathetic eruption of two filaments occurred in tripolar magnetic field region on 2013 October 20. The coronal magnetic field of the eruption source region was a fan-spine topology which hosted the two filaments respectively in its two lobes of the fan. The initiation of the sympathetic filament eruptions was started from the formation of an unstable small filament (F1) through the magnetic reconnection between two crossing mini-filaments (MF1 and MF2). The eruption of F1 was a failed filament eruption, and its start time was coincidence with the sudden emergence of negative magnetic flux in the eruption source region. The physical reason for the failed eruption of F1 was possibly due to the low decay index ($\rm n~\textless~1.5$) above the filament as what had been revealed by the extrapolated coronal magnetic field. During the rising of F1, a loop structure riding on F1 showed interesting writhing and kinking motions before its eruption, which was possibly due to the increased magnetic twist transferred from F1 via the magnetic reconnection between the loop and the rising F1. This process can significantly reduce the magnetic twist or non-potential magnetic energy of F1. Therefore, we propose that the effective reduction in magnetic twist could be a possible reason for the failed eruption of filaments, because there is no enough energy to power the eruptions.

Although F1 failed to erupt in the interplanetary space, it destroyed the equilibrium of the small fan-spine system and F2 hosted by the south lobe of the fan structure. How did the eruption of F1 result in the subsequent eruptions of the small fan-spine system and F2 has been explained in detail in our cartoon, the topology change around F1's eruption source region led to the instability of the nearby magnetic system in which the small fan-spine and F2 resided in. It should be pointed out that the first expansion and then contraction motions observed during the rising of F1 indicated the occurrence of magnetic implosion process, as what has been reported in previous studies \cite[e.g.,][]{lw2010,gos12,liuliu12,lw2010,gos12,Sun12,sim13,rus15}. Here, we propose that the magnetic topology change can also be used as the physical linkage for sympathetic filament eruptions in tripolar fan-spine magnetic system, and the topology change may be led by the magnetic implosion. For the present event, the loss of equilibrium of the south lobe of the fan structure firstly resulted in the null point reconnection and eruption of the small fan-spine system, and then the rising and eruption of F2. The energy releasing around the null point and the eruption of the small fan-spine system can further change the topology above F2. Therefore, the eruption of F2 occurred after the eruption of the small fan-spine system. The eruption of F2 was a successful eruption which caused a large-scale CME in the outer corona. We noted that the start time of F2's eruption was a coincidence with the beginning of magnetic flux cancellation in the photosphere, and the successful eruption of F2 was possibly due to the relatively large decay index ($\rm n~\textgreater~1.5$) of the overlying coronal magnetic field.

Sympathetic filament eruptions have been studied intensively in recent years \citep[e.g.,][]{peng07,shen12a,2020ApJ...892...79S,hou20}, and those occurred in breakout magnetic systems and pseudostreamers often started from external reconnection around the null point due to external or internal disturbances \citep[e.g.,][]{moo01,peng07,shen12a,tor11,Lynch13}. In principle, if the external reconnection starts first, it will greatly consume the overlying confining magnetic field lines and decrease the magnetic tension force upon the low-lying core magnetic structure such as a filament. The consequence of the external reconnection will result in the eruption of the core magnetic structures and the occurrence of the internal reconnection. For the present event, we propose that the external magnetic reconnections occurred after the internal ones during the eruption periods of the two filaments, as the scenario described in \nfig{fig9}. This is mainly based on the observations of the appearance times of the flare ribbons and PFLs. During the eruption of F1, the conjugated flare ribbons of R1 and R2 (PFL1) appeared about 2 minutes before R3 (PFL2). According to the standard filament eruption model within the framework of fan-spine magnetic topology, R1, R2, and PFL1 should be caused by the internal reconnection below the rising filament, while R3 and PFL2 were associated with the external reconnection around the null point. The eruption of F2 showed the same observational characteristics as the eruption of F1, so it also suggested that the internal reconnection below the rising F2 occurred before the external reconnection around the null point of the fan-spine system. 

It should be pointed out that the cartoon shown in \nfig{fig9} can also be used to explain such events in which the external reconnection starts firstly, if we make some minor changes. In this case, since the external reconnections start firstly before the internal ones, the appearance times of the flare ribbons of R1, R2 and R3 should be observed simultaneously during the initial rising phases of the two filaments. For the appearance times of the PFLs, one can expect the firstly show up of PFL2 (PFL1) during the eruption of F1 (F2). Although the eruption of F1 was a failed eruption in the present event, the cartoon also implies the possibility for producing sympathetic CMEs when sympathetic filament eruptions occurred in tripolar fan-spine magnetic systems. In such cases, all the filament eruptions should be powerful enough, or the decay indexes of their overlying coronal magnetic fields are all higher than the critical value for torus instability. In addition, the quadripolar breakout magnetic systems can also launch sympathetic CMEs evolving from sympathetic filament eruptions \citep{shen12a}.

Generally, the fan-spine magnetic system represents the 3D magnetic topology of straight anemone type solar jets \cite{shen21}, and many observational and numerical simulation works are all taken such a special magnetic system as the basic coronal magnetic environment of solar jets \citep[e.g.,][]{par09,str15,wyp18,shen19b,hongj19}. Many recent high spatiotemporal resolution observations showed that solar jets are driven by mini-filament eruptions in fan-spine systems and in association with photospheric magnetic flux cancellations, and these features are also frequently observed in large-scale energetic solar eruptions \citep[e.g.,][]{shen12b,shen17,str15,2016ApJ...830...60H,2017ApJ...835...35H,2017Natur.544..452W,Panesar17,Panesar18,2018ApJ...864...68S,2019Sci...366..890S}. Therefore, small-scale solar jets probably represent the miniature version of large-scale solar eruptions, and hence that this may hint a possible scale invariance of solar eruptions \citep{2016SSRv..201....1R,shen21}. In the line of this thought, the present event also showed some common characteristics with solar jets. For example, the eruption included the eruption of filaments in fan-spine magnetic system, and the start of its main eruption was a coincidence with photospheric magnetic flux cancellations. Typically, solar jets are often associated with small-scale eruption source regions. Therefore, based on the observing capacity of our current telescopes, it is hard to distinguish as many  eruption features as possible to diagnose the formation mechanism of solar jets. The present event occurred in a relatively larger eruption source region, it exhibited more observable clues about its triggering and evolution details. If the  scale invariance of solar eruptions really exists, we can take the present event as a solar jet in a relatively larger version. Hence, the eruption mechanism of the present event can also be used to explain the eruption of solar jets. More observational and theoretical works are desirable in the future to verify our scenario discussed in this paper.

The authors would like to thank the {\em SDO} and other data providing teams for their excellent and user-friendly observations, and we really appreciate the reviewer in reviewing our manuscript with many valuable suggestions and comments. Mr. C. Zhou thanks the helpful discussions with Dr. C. Xia from Yunnan University. This work is supported by the Natural Science Foundation of China (12173083,11922307,11773068,11633008), the Yunnan Science Foundation for Distinguished Young Scholars (202101AV070004), the Yunnan Science Foundation (2017FB006), the National Key R\&D Program of China (2019YFA0405000), the Specialized Research Fund for State Key Laboratories, the Open Research Program of CAS Key Laboratory of Solar Activity (KLSA202017), and the West Light Foundation of Chinese Academy of Sciences.

\end{document}